\documentclass[%
 reprint,
superscriptaddress,
 amsmath,amssymb,
 aps,
pra,
]{revtex4-2}
\renewcommand{\selectlanguage}[1]{}

\usepackage{graphicx}% Include figure files

\usepackage{dcolumn}% Align table columns on decimal point
\usepackage{bm}% bold math
\usepackage{hyperref}% add hypertext capabilities
\usepackage{physics}
\usepackage{quantikz}
\usepackage{mathtools}

\begin{document}
\preprint{APS/123-QED}

\title{The foundational value of quantum computing for classical fluids}

\author{Sauro Succi}
\affiliation{
Fondazione Istituto Italiano di Tecnologia 
 Roma, Italy}%
\affiliation{Physics Department, Tufts University - Somerville, USA}%

\author{Claudio Sanavio}
\affiliation{
Fondazione Istituto Italiano di Tecnologia 
 Roma, Italy}%

\author{Peter Love}
\affiliation{Physics Department, Tufts University - Somerville, USA}%

\begin{abstract}
    
Quantum algorithms for classical physics problems expose new patterns of 
quantum information flow as compared to the many-body Schr\"{o}dinger equation. 
As a result, besides their potential practical applications, they also offer 
a valuable theoretical and computational framework to elucidate the foundations 
of quantum mechanics, particularly the validity of the many-body Schr\"{o}dinger 
equation in the limit of large number of particles, on the order of the Avogadro number. 
This idea is illustrated by means of a concrete example, the Block-Encoded Carleman 
embedding of the Lattice Boltzmann formulation of fluid dynamics (CLB). 
\end{abstract}

\maketitle

\section{Introduction}

In his Nobel speech, Walter Kohn argued that the N-body Schr\"{o}dinger equation (NBSE) 
is unlikely to bear any physical meaning beyond $N \sim 100$ \cite{KOHN}. 
The statement stems from the exponential amount of information contained in 
the $N$-body Hilbert space, in a $d$-dimensional grid with $g$ collocation points per 
dimension, the number of degrees of freedom  scales like 
$g^{dN}$, the usual  \textit{curse of dimensionality} problem. 
This observation was made precise in by Poulin {\em et al.}, who showed 
that physical time evolution can only explore a tiny fraction of the 
available Hilbert space~\cite{POULIN}. 

Kohn and Poulin's arguments raise a far-reaching question for quantum 
information science, namely whether the flow of quantum information in macroscopic 
systems, with $N$ of the order of the Avogadro number, can be organized according to 
patterns other than the NBSE. In this Letter, it is argued that quantum computing 
for classical systems provides a concrete framework to seek operational answers to this basic question. 

As famously proclaimed by Feynman in his trailblazing 1982 paper \cite{FEY}, Nature 
isn't classical, hence if we wish to simulate Nature, we'd better make 
it on quantum computers. Feynman was less explicit on the fact that 
even though Nature is quantum, it has a nearly unstoppable built-in tendency to 
become classical at sufficiently large scale and/or high temperatures.
He implicitly recognized this by adding that the problem is interesting 
because it is not easy at all, classicalization being precisely the reason which
makes quantum computing so hard to realize in practice.
The fight against classicalization through noise mitigation and quantum error 
correction is a mainstay of current quantum computing research, but in this Perspective 
we address a different question, namely whether a classical system can
be simulated according to quantum mechanical rules (not necessarily NBSE) 
and possibly faster than on a classical computer.

In principle the first part of this question may seem circular; given that the world is quantum 
mechanical, and classical physics emerges from quantum mechanics, at the 
fundamental level, any classical system must be implicitly computed quantum mechanically.   
This assumes that Nature can afford the luxury of computing all the way in Hilbert
space, wasting most of its marbles on empty regions, while classical physics
emerges through non-unitary dynamics such as decoherence and/or measurement. 
On closer inspection, however, the question is far from empty, because there
might exist quantum algorithms for the simulation of classical physics
which are {\em not} based on the emergence of classicality from the NBSE. 
Assessing the existence of such algorithms is conceptually important because it 
offers concrete alternatives to the  $N$-body Schr\"{o}dinger equation for
macroscopic matter, thereby putting flesh into Kohn's speculations.   
Whether they can outdo their classical counterparts is 
a separate and much more difficult question, which we also
address in this Letter. 

We investigate these matters by means of a concrete example, the formulation
of a quantum Carleman-Lattice Boltzmann algorithm for classical fluids. 
Before dwelling into the details of this specific approach, let us summarize 
the main guidelines of the general framework such specific method belongs to.

{\it i) Discretization}: we ultimately aim at concrete quantum
simulations, hence we consistently deal with large but finite numbers of degrees of freedom. 

{\it ii) Uplifting}: classical systems are typically dissipative, meaning that 
they leak information to the surrounding environment in an irreversible way. 
Reversibility can be restored by enlarging the state space so as to include 
extra-degrees of freedom (environment) absorbing the
information lost by the system, so that the System+Environment (Universe) 
is reversible and can consequently be described by a unitary dynamics. 

{\it iii) Linear Embedding}: classical systems are most often nonlinear, hence incompatible
with the quantum superposition principle which lies at the roots of quantum computing.
The nonlinearity can be traded for extra-dimensions via linear embedding of the 
dynamics into infinite-dimensional spaces and then truncated to a finite order (see point 1).

{\it iv) Nonlinear Depletion}:
The truncation order is strictly related to the strength of the nonlinearity, 
whence the scope for formulations which present the least nonlinear strength.
Several techniques are known in theoretical physics to weaken the nonlinear
coupling, renormalization group techniques and AdS-CFT duality being two prominent examples
in point \cite{RG,ADSCFT}. In our case, this is achieved by simply moving 
to a phase-space representation of the system, i.e. the Boltzmann kinetic 
level \cite{BSV}. 

{\it v) Quantum Simulation}: the final goal of the program is not only
to provide complexity estimates but to deliver a concrete quantum algorithm 
and associated quantum circuit to be simulated on actual quantum hardware.

Progress on all these steps {\it i}-{\it iv} has been made in a number of disconnected works. Reversible microscopic models such as the hydrodynamic lattice-gas automata are a form of both discretization and uplifting - replacing irreversible nonlinear fluid dynamics by reversible nonlinear dynamics of particles on a lattice~\cite{FHP}.  Other approaches to uplifting fluid equations to make them Hamiltonian, and hence reversible, have been given in~\cite{Becker}. Linearization is the basis for many current quantum approaches to nonlinear differential equations, including Carleman approaches~\cite{CHILDS,CHEUNG} discussed here but also ``replica'' methods based on linear evolution of many copies of the system~\cite{seth}. Renormalization techniques for partial differential equations are well established~\cite{RG,Barenblatt} and of current interest~\cite{bubble}. Techniques for reducing nonlinearities in field theories by similarity renormalization have also been applied to fluid equations in~\cite{Jones1}.  In this article we argue that quantum algorithms for fluid dynamics require further progress in all steps, and raises interesting foundational questions about the emergence of nonlinearity and the transition to classicality. Having clarified the conceptual framework, a few general comments on quantum computing for fluids are now in order. 

\section{Quantum computing for fluids}

Quantum computing emanates from two basic properties of quantum mechanics:
{\it Linearity} and {\it Unitarity}~\cite{Deutsch1985,NIE}. 
The physics of fluids is generally neither, hence
the solution of the fluid equations on quantum computers immediately 
faces two major obstacles: Nonlinearity and Dissipation
\cite{Succi_EPL}.  
In the following we present one out of many possible strategies around both obstacles: Carleman 
embedding combined with block-encoding of sparse matrices.

\subsection{Dealing with non-linearity: Carleman embedding}

Carleman embedding is  an uplifting technique whereby a finite-dimensional
non linear system is formally turned into an infinite-dimensional 
linear one \cite{CARLE}.
Hence, the basic idea is to trade nonlinearity for infinite-dimensionality. 

\begin{figure}[h]
  \centering
  \includegraphics[width=0.5\textwidth]{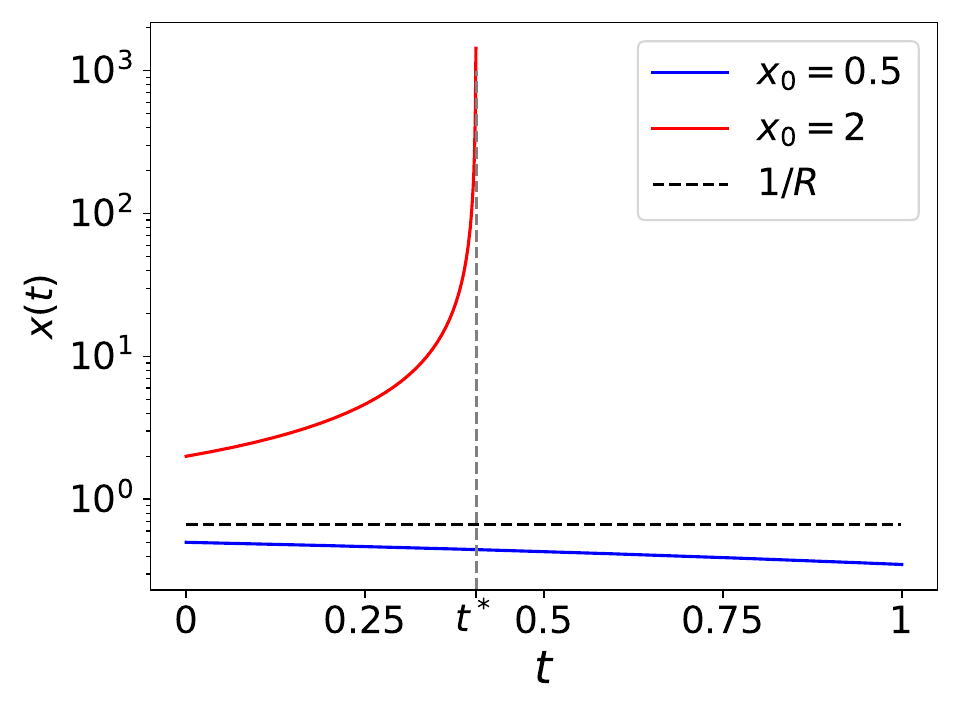}
  \caption{The converging (blue solid line) and the diverging (red solid line) solutions of the logistic equation, obtained by setting the initial condition smaller, $x_0 = 0.5$, and larger, $x_0=2$, than $1/R$ respectively, with $R=1.5$ (black dashed horizontal line). The gray vertical line marks the time singularity for the unstable solution $t^*$.}
  \label{fig:carleman_logistic_a}
\end{figure}

As an elementary and yet representative example, let us consider
the logistic equation
\begin{equation}
\label{LOGI}
\dot x = -ax + bx^2
\end{equation}
with initial condition $x(0)=x_0$ and $a,b>0$.
The coefficient $a$ mimics dissipation while the ratio $R=b/a$ measures
the strength of the quadratic nonlinearity versus dissipation, the "analogue"
of the Reynolds number in actual fluids (which also feature a quadratic nonlinearity).
The logistic dynamics shows two attractors, a stable one at $x_S=0$ and
an unstable one at $x_U=1/R$. This means that any initial condition below
$1/R$ decays asymptotically to zero, while for $x_0>1/R$, the solution 
exhibits a finite-time singularity at $t=t^*=a^{-1}\log(\frac{r}{r-1})$, 
as it is apparent from the analytical solution
$$
x(t) = x_0 \frac{e^{-at}}{1-r + r e^{-at}}
$$ 
where we have set $r = Rx_0$.
Initially, this solution decays as $e^{-at}$,
as long as $|r (e^{-at}-1)| \ll 1$, followed by a slower decay
but still converging to zero as $\frac{e^{-at}}{1-r}$, provided $r<1$, 
see Fig.~\ref{fig:carleman_logistic_a}.

%---------------------------------------------

Carleman embedding sets out to capture this multi-timescale
relaxation through the progressive insertion of extra-variables, each
describing an increasingly longer time scale.

\begin{figure}[h]
\centering
\includegraphics[width=0.5\textwidth]{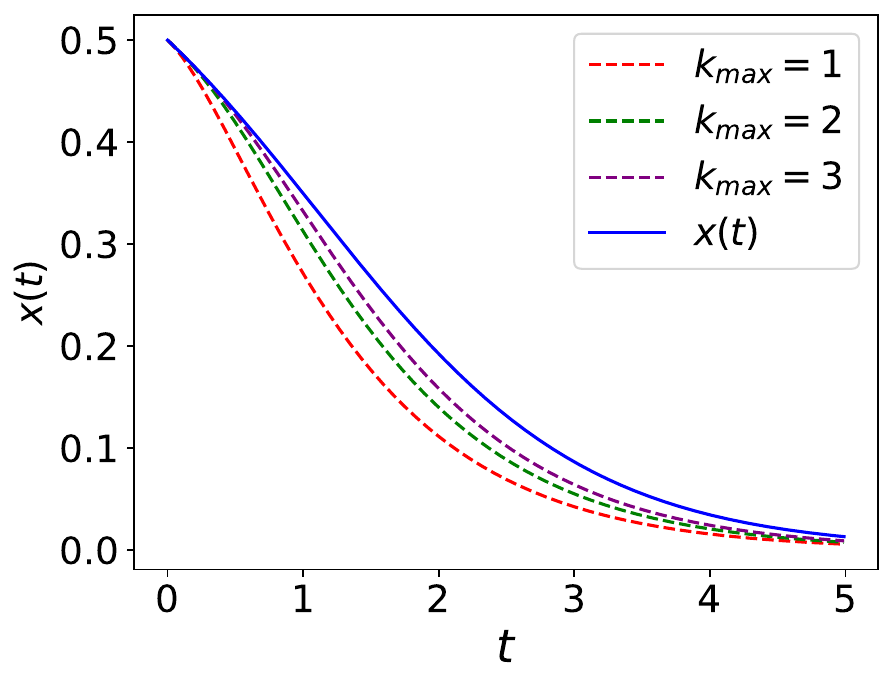}
\caption{The analytical solution of the logistic equation in the stable regime (blue solid line) is compared 
with the solutions of the Carleman system of equations with increasing truncation order $k_{max}$, which 
yields a better approximation of the solution. We set $x_0=0.5$ and $R=1.5$, hence $r=3/4$.}
  \label{fig:carleman_logistic_b}
\end{figure}

In practice, upon letting $x_1 \equiv x$ and $x_2 \equiv x^2$, the  
logistic equation rewrites as
$$
\dot x_1 = -ax_1 + bx_2
$$
This is linear, but open, since $x_2$ is formally a new unknown.
The equation for $x_2$ is readily derived 
$$
\dot x_2 = 2 x \dot x = -2ax^2 + 2bx^3 \equiv -2(ax_2-bx_3)
$$
The name of the game is quite clear, the embedding 
$$
x \to \lbrace x_k \equiv x^k \rbrace,\;k=1,k_{max}
$$
turns the original nonlinear problem into an infinite hierarchy
$$
\dot x_k = -k (ax_k - b x_{k+1})
$$
Occasionally, this linear hierarchy of ODE's can be integrated 
analytically in the limit $k_{max} \to \infty$, thereby recovering
the exact solution.
More typically, the hierarchy is truncated at a given level $k_{max}$
by setting $x_{k_{max}+1}=0$, thereby providing a closed approximated solution
to the nonlinear problem. In Fig.~\ref{fig:carleman_logistic_b} we show the 
approximate curves at increasing values of $k_{max}$ for the stable solution with $x_0<1/R.$ 

% -------------------------------------------------------

\begin{figure}[h]
  \centering
  \includegraphics[width=0.5\textwidth]{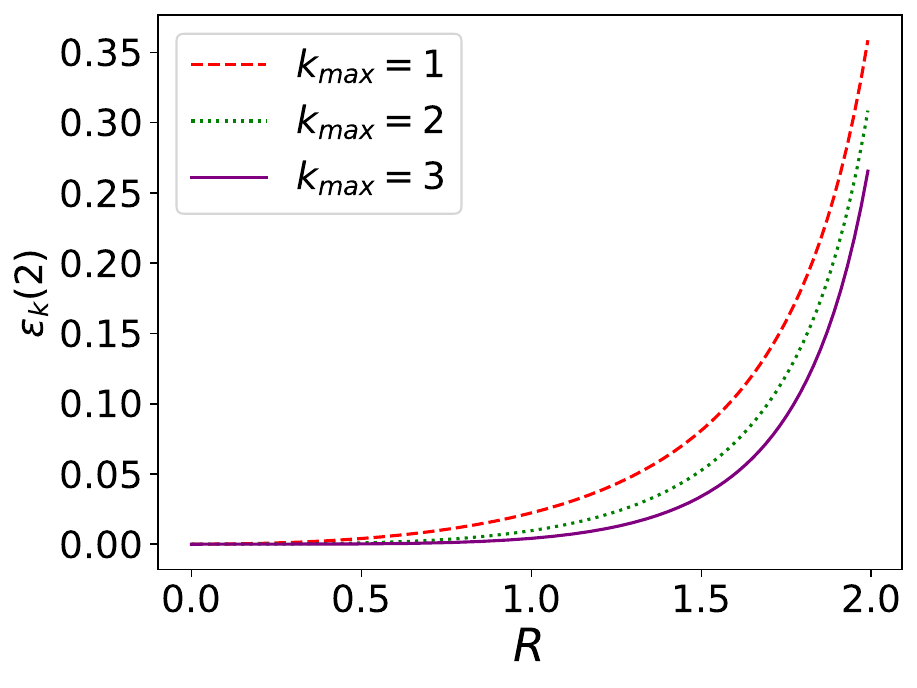}
  \caption{The accuracy $\epsilon_k(t)$ calculated at $t=2$ for given $k_{max}=1$ (red dashed line), 
  $k_{max}=2$ (green dotted line) and $k_{max}=3$ (purple solid line). 
The  initial condition is set to $x_0=0.5$, therefore 
 the stable region is $0<R<2$.}
  \label{fig:carleman_logistic_c}
\end{figure}

The basic idea is that low order truncations may offer cheap approximations
within a finite time interval $0<t<t_{max}$.
On intuitive grounds, one expects that at a given level 
of accuracy $\epsilon_k(t) = |x(t)-x_k(t)|$ at given time $t$ and for 
a defined truncation cutoff $k_{max}$, the latter should
be an increasing function of $R$. The specific shape of this function 
dictates whether or not trading non-linearity for higher-dimensionality
is a good bargain.
In Figure~\ref{fig:carleman_logistic_c} we show how the 
error $\epsilon_k(t)$ changes with the non-linearity parameter $R$. 
The same level for $\epsilon$ is reached at larger $k_{max}$ when $R$ increases.

% --------------------------------------------------------------- 
\begin{figure}[h]
  \centering
  \includegraphics[width=0.5\textwidth]{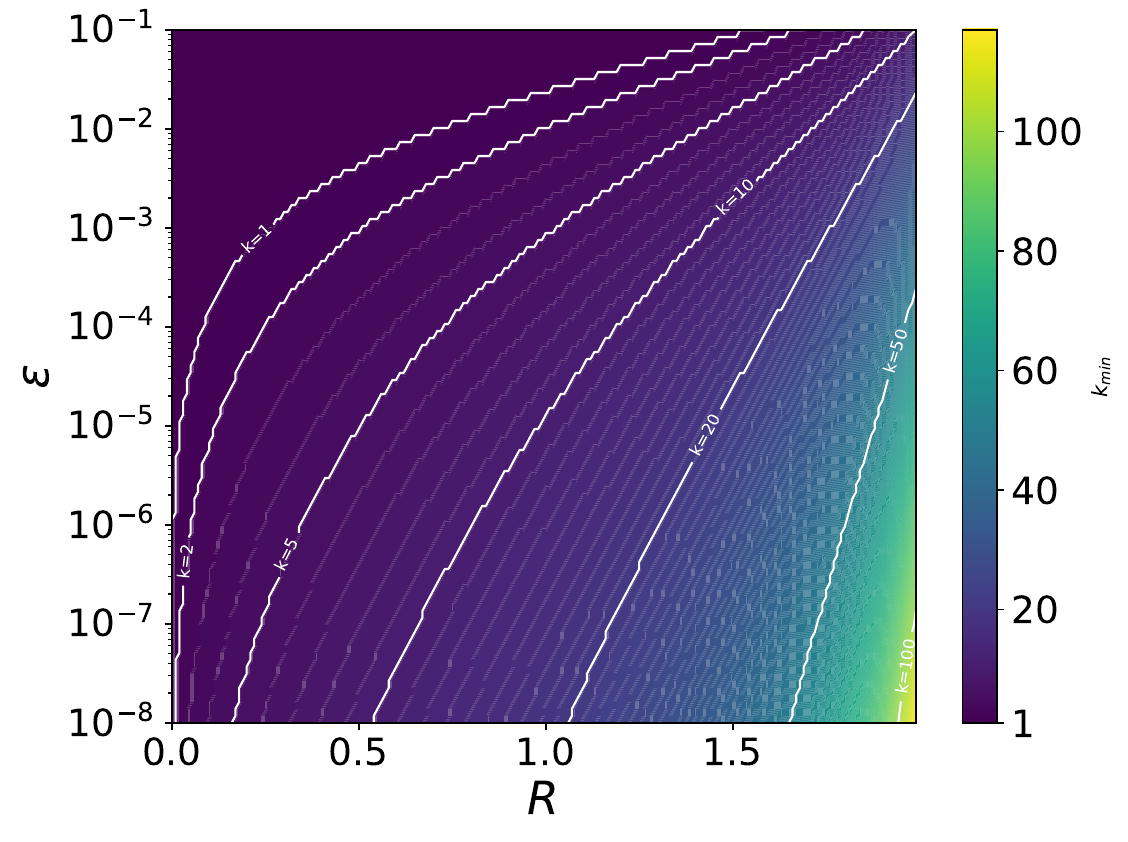}
  \caption{The minimum value $k$ of the truncation order $k_{max}$ to achieve accuracy 
  $\epsilon$ for given nonlinearity $R$. $\epsilon$ is calculated at time $t=2$ and 
  represented in log scale, while the initial condition is set to $x_0=0.5$ and defines the stable region 
  $0<R<2$. The isolines for $k$ are drawn in white.}
  \label{fig:carleman_logistic_d_isolines}
\end{figure}

This is made clear in Fig.~\ref{fig:carleman_logistic_d_isolines}, where we show the minimum 
cutoff $k_{min}(R,\epsilon)$ that one should use to solve the logistic equation with fixed 
accuracy, for a given nonlinearity $R.$ 

% ---------------------------------

On classical computers, Carleman linearization can be solved by a number of 
techniques, but does not seem to have gained any prominent role. 
On quantum computers, it provides an elegant and appealing strategy 
to eliminate nonlinearity.
This does not come for free, since a Carleman scheme on a grid with $G$ grid points 
and $k$ levels of approximation takes about
\begin{equation}
N_C \sim G^k
\end{equation}
variables, hence a very large matrix problem.

Carleman linearization for quantum simulations was first applied to the Burgers equation
with encouraging results \cite{CHILDS}. 
However the Burgers equation, besides being one-dimensional, is also pressure-free
which is a drastic simplification of the physics of fluids.
Subsequent application to the two-dimensional Navier-Stokes 
equations~\cite{sanavio2024a} has shown very poor convergence, mostly 
on account of the non-local coupling between the flow and pressure fields. 
However, once applied to the lattice Boltzmann formulation
\cite{Ita22,sanavio2024b} of fluid dynamics, it has shown extremely 
encouraging results, with errors around $\epsilon \sim 10^{-4}$ for 
hundreds of time-steps even at the
lowest rung of the Carleman ladder, namely $k_{max}=2$.
The reasons for this excellent performance have been discussed 
at length in the original papers, but essentially they amount to the
fact that the nonlinearity in the dynamic of phase-space fluids
is controlled by the Mach number (typically order 1) instead of the
Reynolds number (typically order millions and billions). 
Moreover, the free-streaming operator is exact and unitary.

Unfortunately, these excellent properties are not sufficient to deliver an
efficient quantum algorithm, the main problem being that the Carleman LB
matrix projects upon virtually all of the tensor Pauli basis matrices.
Symbolically, upon expanding the Carleman matrix onto the tensor Pauli basis $P_l$, 
$C_{ij}=\sum_{l=1}^{N_c} c_l P^{(l)}_{ij}$, it is found that the coefficients scale 
like $|c_l| \sim 1/l$.
This means that none of them can be ignored and the depth of the 
resulting quantum circuit scales like $N_c^2 = 2^{2q}$,like a random unitary. 

This discomforting outcome can be circumvented by moving to a sparse-matrix
representation of the Carleman matrix, whereas each non-zero element
$c_{ij}$ is representd by two types of oracles, one providing the locations
$j=1,2 \dots s$ ($s$ is the sparsity of the Carleman matrix) such that $c_{ij} \ne 0$, and 
the other providing the non-zero values $c_{ij}$ themselves.
The number of ancilla qubits is fixed by the sparsity of the Carleman matrix
$q_a \sim log_2 s$.
The explicit form of these oracles has been worked out and shown to bring the
exponential depth down to a quadratic one.   
The interested reader is kindly directed to the original literature 
\cite{CHEUNG,sanavio2024b,sanavio2025}.

Yet, this leaves us with another problem, namely the fact that the implementation
of the oracles requires extra qubits, known as ancillas, which in turn imply a
non-zero failure rate of the quantum update, an issue that we are going
to discuss in the next section.

\subsection{Dealing with dissipation: Block Encoding}

Next, we address the second obstacle: non-unitarity.
A number of strategies are available to turn a dissipative system
into a conservative one, but the most popular one is the so 
called {\it Block-Encoding} (BE), whereby the quantum system mapped into the
state $|\psi\rangle$ of $q_s$ qubits is augmented with a number $q_a$ of 
auxiliary qubits known as "ancillas" $|a\rangle$, representing the environment
\cite{CAAMPS}. 
The enlarged state $|\Phi\rangle \equiv |\psi\rangle|a\rangle$ is acted upon
by a correspondingly augmented Carleman operator $\hat C_{BE}$, such that
the BE update formally reads as follows:
\begin{equation}
\label{BECLB}
\dot f = C_{BE} f
%\frac{d |\Phi\rangle}{dt} = \hat C_{BE} |\Phi\rangle
\end{equation}
where $f$ is the set of Carleman variables at a given truncation level $k$.
For instance in the case $k=2$ this is the set of one and two-body
distributions $f = \lbrace f_i(x_1), f_{ij}(x_1,x_2) \rbrace$,
where $x_1$ and $x_2$ are spatial coordinates and the
subscripts $i,j$ label the discrete velocities. 
The embedding of this classical two-body problem into a quantum
representation is described in the original papers \cite{sanavio2024b}.
%Note that the carleman matrix $C_{BE}$ is the analogue of the 

\begin{figure}
    \centering
    \includegraphics[width=\linewidth]{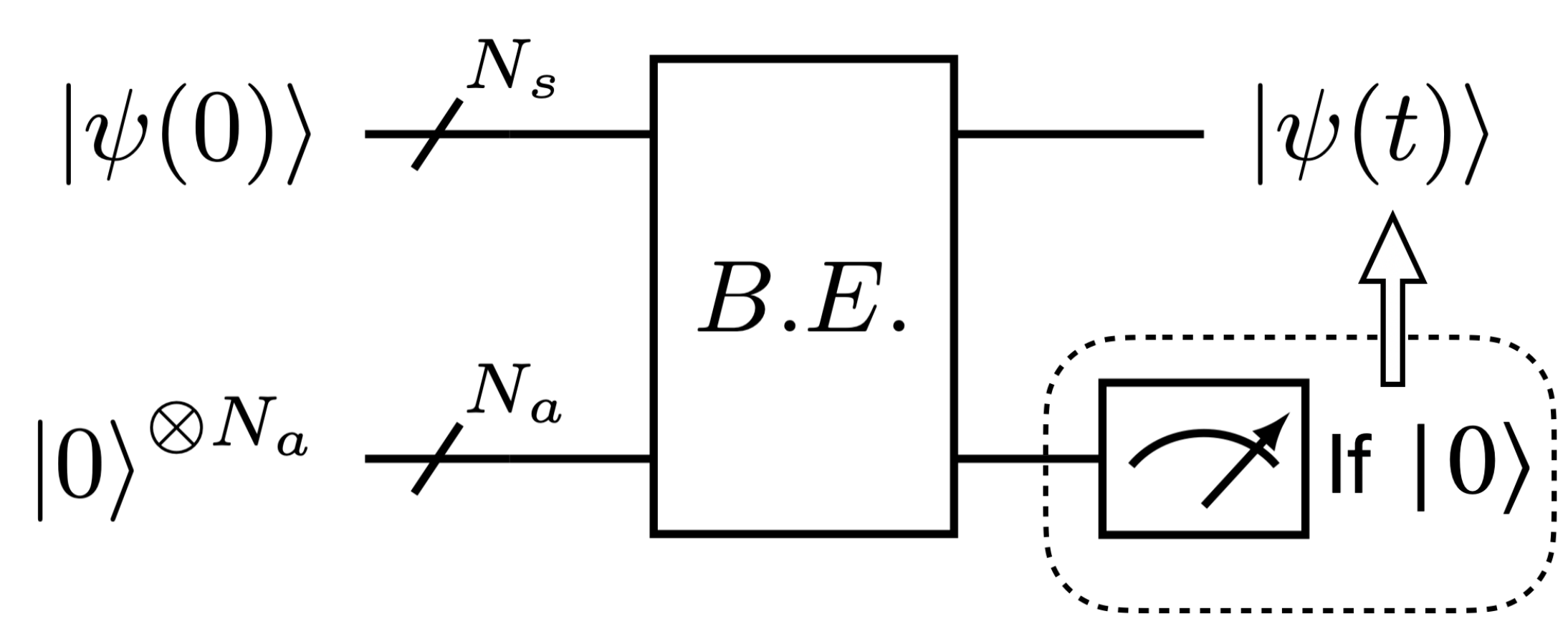}
    \caption{The quantum circuit for block encoding a nonunitary operation into the unitary operator B.E. The success of the algorithm is conditioned on measuring all the $q_a$ ancilla qubits in the state $|0\rangle$. 
This happens with a probability $p \sim 2^{-2q_a}$, with little dependence upon the number of system qubits $q_s.$ }
    \label{fig:circuit_BE}
\end{figure}

A schematic quantum circuit is shown in Fig~\ref{fig:circuit_BE}.
It is easy to show that this update recovers the original and correct one
only whenever all the ancilla qubits are aligned in state $|0\rangle$, in 
which case they do not "contaminate" the update. 
The upper bound for this to happen can be estimated as 
$p_a \sim 2^{-2q_a}$, highlighting a severe constraint at increasing 
number of ancilla qubit..

% -----------------------------
% --------------------------

Block-Encoded Carleman-LB (BECLB) algorithms have been developed in the
last few years, the bottom line being that the corresponding circuits 
offer a favorable (quadratic) scaling of the circuit depth with the
number of qubits~\cite{sanavio2025}. However, the success probability of the dissipative update
is pretty low, of the order of $10^{-4}$ for a single time-step, thus
compromising the viability of multi-step integration.
A possible way out is offered by {\it telescopic} quantum algorithms, whereby the
solution at a given finite time $t$ is reached within only a few
timesteps, ideally just one, so as to curb the effects of low success probability.
Again, the detail-thirsty reader is kindly directed to the 
original literature \cite{WIEBE,Zecchi2025b}.  

Next we reconnect with the main theme of this paper, namely the 
foundational value of quantum algorithms as "alternatives" to the
NBSE equation in the limit $N \to \mathcal{A}$, the Avogadro number.

\section{Quantum lessons from the water faucet}

For the sake of concreteness we refer to the simulation of 
the very ordinary case of fluid physics: the water flow 
from a kitchen faucet. The reason is that such flow is within grasp
of current BECLB algorithms \cite{sanavio2024b}.

Let us recall that the number of active dynamic degrees of freedom ("eddies")
in a fluid over a time span $t \sim L/U \sim G^{1/3}$ is given by \cite{FRISCH}:
$$
N_{dof} = Re^3
$$
where $Re=UL/\nu$ is the Reynolds number, $U$ being the flow speed,
$L$ a typical macroscale and $\nu$ is the kinematic viscosity.
For a water faucet, $U=1$ ($m/s$), $L=0.01$ ($m$) and 
$\nu =10^{-6}$ ($m^2/s$), so that $Re \sim 10^4$ and $G=10^9$. 
The number of floating point operations required
to complete a dynamic simulation over a time span $t \sim L/U$ is approximately
$10^3 Re^3 \sim 10^{15}$, meaning that a mid-end Teraflops/s computer can
simulate this flow in about one hour wall clock time.
This is where classical computing for classical fluids (CC) stands today \cite{CFD}. 

Next, let us address the other extreme, the same flow taken head-on via the NBSE.
A centimeter cube of water contains about $10^{22}$ molecules,
which we equate for simplicity to the Avogadro number 
$\mathcal{A} \sim 6 \times 10^{23}$.
This is basically the number of dimensions of Hilbert space 
underneath the humble faucet flow.      
With a modest $g=10$ grid points per dimension, this leads to
$G_{NBSE} \sim g^{\mathcal{A}}$ grid points, to be compared
with the $G_{NS}= Re^{9/4} = 10^9$ grid points required by the
solution of the Navier-Stokes equations on a classical computer:
Kohn's point in full glory.

How about quantum computers? Assuming a perfect logarithmic scaling, we 
are left with the order of $\mathcal{A}$ qubits, still completely 
undoable for any foreseeable quantum computer and astronomically 
more costly than the classical simulation.

The question comes back again: Nature is hierarchical and modular, in 
that at each level offers "effective" descriptions based 
on the relevant degrees of freedom of that specific level. 

In our case, a single fluid degree of freedom ("eddy")
contains about $\mathcal{A}/G_{NS}=10^{14}$ molecules, this is the information
compression associated with the passage from the Schroedinger 
to the Navier-Stokes levels.

If we insist that Nature, being quantum, must necessarily compute according to
quantum mechanics, we come to the rather puzzling conclusion 
that Nature, as an analogue quantum computer, ignores
the perks offered by its own "emergent" properties.

Three options then arise, one bad, one good and one golden.

1) {\it The bad}: Nature is a super-powerful analogue quantum 
computer and as a such, it can afford the luxury of computing 
quantum mechanically all the way, according to the NBSE. 
If so, Kohn's argument does not apply: we can't compute with NBSE but Nature can.

2) {\it The good}: There are ways of computing macroscopic classical systems 
according to quantum mechanics (linear and unitary) on different and more 
economic grounds than the NBSE, yet less efficient than classical fluid solver. 
This has significant foundational value because it points to
concrete alternatives to the NBSE, e.g. in our case the equation (\ref{BECLB}). 
Kohn is right, but this has no practical impact on the 
use of quantum computers for fluids.

3) {\it The golden}: Quantum algorithms are not only faster than the 
NBSE but also than classical computers. Besides the foundational 
value, this would mark a major practical breakthrough. 

It is easy to see that the positioning of CLB is crucially dependent
on the single-step success probability.
A CLB$^2$ (CLB truncated at second order) features about $G^2$ Carleman
variables, hence $10^{18}$ for the water faucet, corresponding to approximately
$18 log_2 10 \sim 60$ {\it physical} qubits in an ideal quantum-computing 
world with perfect error-correction algorithms.
  
This is astronomically less than the Avogadro-like number of qubits required by NBSE.
However, with probability of success of the order of $10^{-4}$ a $10^3$ step
simulation succeds in reproducing the correct quantum state with a probability
$10^{-4000}$, which completely defeats the purpose of CLB 
(making it much worst than NBSE). 
%(note that
%$10^{4000} \gg \mathcal{A}$, hence CLB is far more expensive than NBSE) . 

As mentioned earlier on, a possible way out is to develop telescopic algorithms 
capable of reaching the final state in a handful of time-steps, ideally
just one. Roughly speaking, to bridge the gap with 
classical computing (on the optimistic assumption that the quantum clock ticks
at the same rate as classical ones), one would need to meet the following condition:
$60$ qubits with $p$ wins over $10^9$ grid sites provided
$$60/p<10^9.$$ 
This means one step with $p_1=60/10^9 \sim 10^{-7}$, 
two steps with $p_1= 10^{-3.5}$ and so on.

More generally, a BECLB update with $k$ Carleman levels, $T$
timesteps and a single-step success probability $p$, wins over the 
classical simulation on a grid with $G$ grid ponts, provided
$
p^T > \frac{k log G}{G},
$ namely:
\begin{equation}
p > p_{min}(G,T;k)= \left(\frac{k log G}{G}\right )^{1/T}
\end{equation}
The value of the minimum success probability (MSP) 
$p_{min}$ as a function of $G$ and $T$ is plotted 
in Figure 6 for the case $k=2$.
From this figure it is clear that in order to compete with
classical simulations, the BECLB procedure must feature
nearly perfect success probabilities, unless an extremely  efficient
telescopic version can be developed.
In the latter case, a ten-step telescopic algorithm would
be competitive at a comparatively low success probability
$p_{min} \sim 0.18$, three orders of magnitude above the
current values. And the ideal case, $T=1$, would work
with $p_{min} \sim 10^{-7}$.

\section{From water faucets to numerical weather forecast}

The previous section dealt with the "mundane" case of the water
faucet because the corresponding Reynolds number is in the 
range of what can be achieved by emulating CLB on present-day 
classical computers. However, the real breakthrough would be to
perform fluid simulations which exceed the capabilities of foreseeable
classical computers. A prominent example in point is numerical weather
forecast \cite{PALMER}. The Reynolds number associated to the
global atmospheric circulation is of the order of $Re \sim 10^{12}$,
which implies $G \sim 10^{27}$ grid points and roughly $10^{40}$
floating point operations for a simulation of 
$T \sim G^{1/3} = 10^9$ time-steps. 
These numbers speak for themselves as to the impossibility 
of any foreseeable classical computer to come any near to such target.

On the {\it strong} assumption that CLB can still compute within
a few Carleman iterates because the nonlinearity is controlled
by Mach and not Reynolds, the MSP for such calculation is
$
p_{min}(G=10^{27},T=10^9,k=2) \sim 0.99999994 
$
namely a maximum failure rate $f_{max}=1-p_{min} \sim 5 \;10^{-8}$, i.e. 
50 parts per billion.
This is extremely small, and yet higher than the failure 
rate of single base DNA replication (after proof-reading and
post-replication mismatch repair), about one in ten billions. 
Hence, even discarding tomographic costs, succesfull quantum computing for 
numerical weather forecast requires genetic accuracy!
Unfortunately, this is not compatible with the constraints set by the
ancillas, $p_{max} \le 2^{-2q_a}$, hence telescopic versions are a must.
 
% --------------------------------------------------
\begin{figure}
\label{FIGMSP}
\includegraphics[scale=0.3]{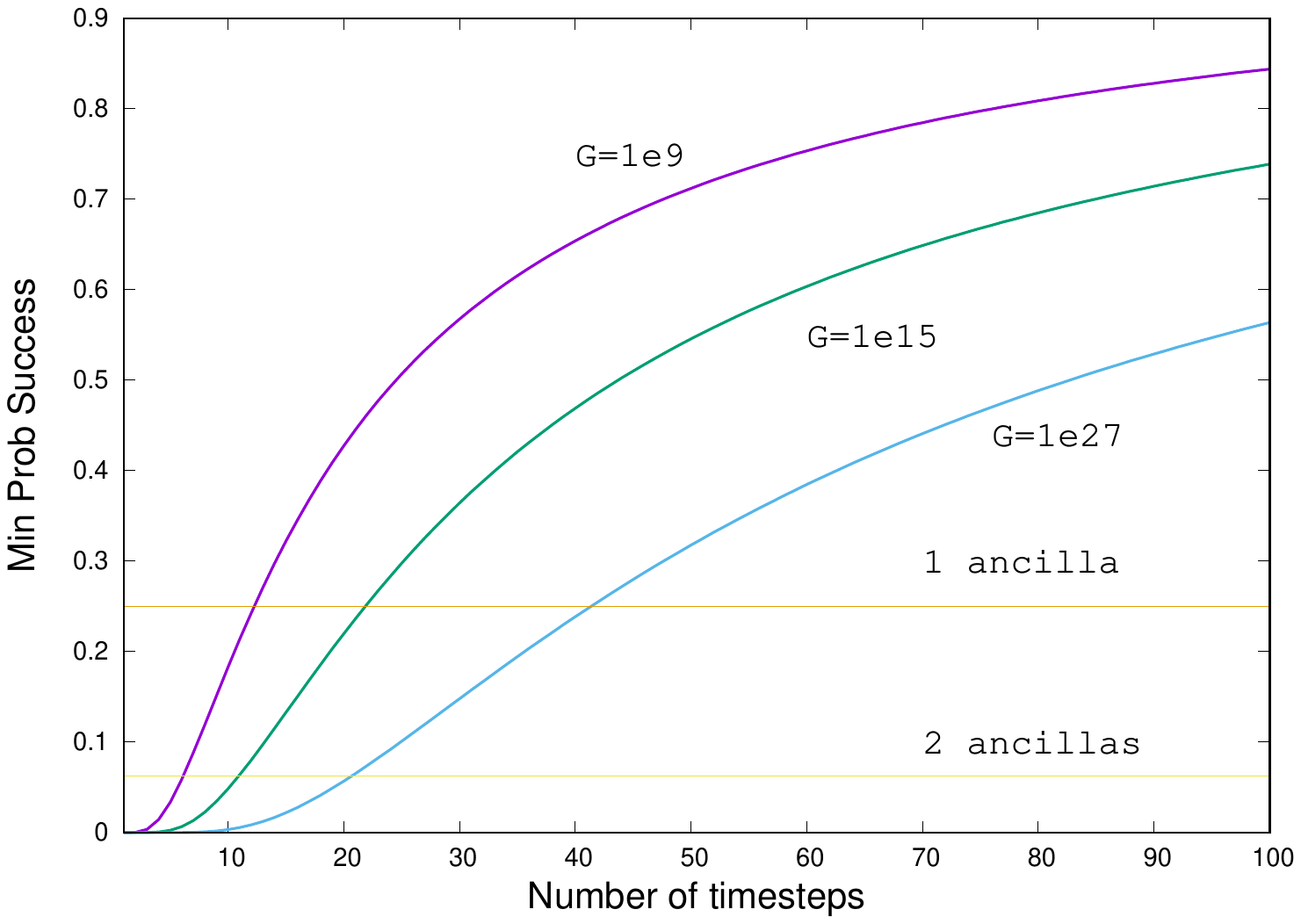}
\caption{The minimum success probability as a function of the 
number of time steps for three representative grid sizes
$G=10^9$ (water faucet), $G=10^{15}$ (full airliner), $G=10^{27}$ (global
weather forecasting. The horizontal bars correspond to the maximum 
value allowed by the number of ancilla qubits.
The crossing between the curves and the horizontal lines defines the
maximum number of timesteps above which quantum advantage is lost
(The figure does not account for tomographic costs of reconstructing the state
after copying to proceed to the next step).
Note that even though the number of time steps $M$ grows with increasing
grid size, the effect is illusory when measured in terms of the classical
number of steps $G^{1/3}$. 
This largely offsets the benefits of the $(k log G)/G$ reduction.  
}
\end{figure}

The figure clearly shows that once the ancilla barrier is accounted for, the
quantum time marching should not employ more than a few ten steps at most.
This is an extremely severe constraint, especially if measured in natural
units of the $G^{1/3}$ steps required by the classical simulation.   
The problem of low success rates is currently being addressed via extensions 
of the oblivious amplitude amplification method to non-unitary matrices
\cite{Zecchi2025a}. 

Summarizing, we have discussed three approaches, N-body schroedinger
(NBSE), Carleman Lattice Boltzmann (CLB) and classical Navier-Stokes (NS).

On a grid with $G$ lattice sites, the three approaches involve
$G^N$, $G^k$ and $G$ degrees of freedom (order of magnitude), respectively.
On quantum computers, these entail $N log_2 G$, $k log_2 G$ qubits, while
NS remains $O(G)$ since this is the classical touchstone.
For a multistep time marching with $T$ timesteps and assuming for simplicity 
the same computational cost per timestep and degree of freedom, we have
$TG^N T$, $TG^k p^{-T}$ and $TG$, where $p$ is the single-step CLB probability
of success and we have assumed that for NBSE such a probability is $1$
because the algorithm is genuinely quantum.
Based on the above, CLB is competitive towards NBSE as long as
\begin{equation}
\label{WIN1}
p \ge G^{\frac{k-N}{T}}
\end{equation}
Since $k<<N$, a grid with $G=10^{12}$ (basically the exascale target) delivers
$p \ge 10^{-12N/T}$, meaning that with $p=0.1$ one can perform $T=12N$ steps,
a pretty long stretch indeed. But the practical point is to outdo classical
NS, hence the condition is:
\begin{equation}
\label{WIN2}
p \ge  (\frac{k log_2 G}{G})^{1/T}
\end{equation}
As discussed earlier on, this extremely more restrictive.

\subsection{Multiscale strategies}

Another possibility, still largely unexplored to the best
of the author's knowledge, is the use of AI tools to learn the
correct form of the telescopic propagator, based upon training  on 
large scale classical fluid dynamics datasets.  
Formally, this goes as follows: consider the time evolution of the 
fine-grained quantum system from time $t=0$ to time $t$:
\begin{equation}
|\psi_t\rangle =  \hat \tau_t |\psi_0\rangle
\end{equation}
where $\hat \tau_t$ is the fine-scale time propagator over a fine grid 
with $G_f$ grid points and $M_f=t/\Delta t_f$ time-steps.
Next, let us project the initial state on a coarser grid with $G_c= (G_f/B^d) \ll G_f$ grid 
points, $B \gg 1$ being the spatial blocking factor along each direction, including time.
Formally:
\begin{equation}
|\Psi_0\rangle =  \mathcal{P} |\psi_0\rangle
\end{equation}
where $\mathcal{P}$ is a suitable projector (Encoder, in machine learning language).
Next evolve the coarse-grained quantum state with a coarser time 
propagator $\hat T_t$, to obtain the coarse-grained state at time $t$:
\begin{equation}
|\Psi_t\rangle =  \hat T_t |\Psi_0\rangle
\end{equation}
%In the simplest instance, the coarse time propagator could just be the same as the
%fine-grain one, just with a correspondingly $B$-times larger timestep.
Finally, reconstruct the fine-grained state at time $t$ via a
reconstruction operator $\mathcal{R}$ (Decoder in machine learning language):
\begin{equation}
|\tilde \Psi_t\rangle =  \mathcal{R} |\Psi_t\rangle = 
\mathcal{R} \hat T_t \mathcal{P} |\psi_0\rangle
\end{equation}

The error introduced by coarse-graining is then given by:
\begin{equation}
|| |\Psi_t\rangle - |\tilde \Psi_t\rangle || =  
||(\hat{\tau}_t - \mathcal{R} \hat T_t \mathcal{P}) \; |\psi_0\rangle ||
\end{equation}

If one could secure that the decoder is exactly 
the inverse of the encoder, $\mathcal{P} \mathcal{R}=\mathcal{R} \mathcal{P} = I$,
no information would be lost in the process, yielding 
approximately a factor $B^4$ saving in computational resources.
This is generally impossible, but machine learning can help
minimize the coarse-graining error described above.
%one would have $\hat T_t = \hat \tau_t$
%hence $|\tilde \psi_t \rangle =  |\psi_t\rangle$: the coarse-grained evolution would be 
%perfect, $B^4$ times cheaper at zero error cost.
%Of course, no such free lunch is possible in actual practice, but one
%can use machine learning to learn the optimal reconstruction
%operator so as to minimize the coarse-graining error.
Such kind of techniques have been recently developed by the computational
fluid community, and shown significant computational savings \cite{KOCH}, typically
one order of magnitude in each spatial dimensions, as well as in time. 
There is no reason why they should not carry to the quantum computing context.
A related variant is to use  quantum-informed machine learning approaches 
as recently proposed for the simulation of high-dimensional chaotic systems\cite{Wang2025}.
Yet another intersting variant of is to use classical dynamics as a coarse-grained solver 
and quantum simulation as a fine-grain solver over a small time-stretch,
idea being that the fine-grain evolution would "heal" the errors incurred by 
the coarse-grained solver (See Fig. 7). 
% -------------------------------------------------
\begin{figure}
\centering
\includegraphics[scale=0.125]{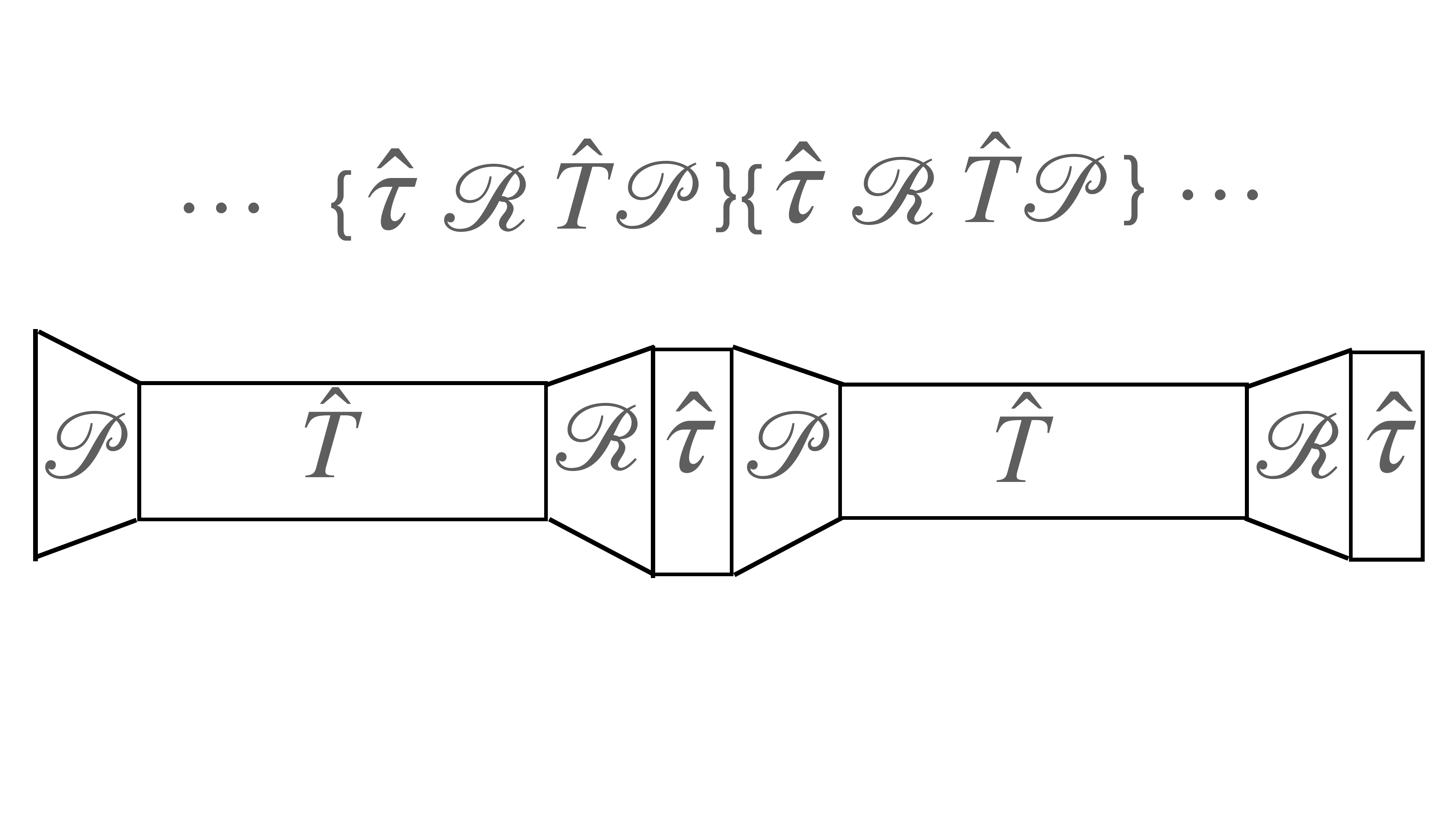}
\caption{Schematics of the multiscale procedure described in the text.
The idea is to run fine-grain quantum simulations for short stretches of time 
$\tau$ with the fine-grain propagator $\hat \tau$) 
so as to curb the effects of low success probabilities,
and perform long stretches of sixe $t$ with coarse grained (classical) solvers ($\hat T$).
Machine learning can help finding optimal versions of the projection and
reconstruction operators minimizing the coarse-graining errors. 
}
\end{figure}

At this stage, it is impossible to predict whether telescopic quantum marchers, possibly
equipped with machine-assisted multiscale coarse-graining, will ever hit
the target of outdoing classical simulations. 
The topic is exciting and up for grabs.

\section{Summary}

Summarizing, we have pointed out by means of a concrete example
that the search for quantum algorithms for fluids bears a 
significant foundational value besides the potentially practical one.
However, realizing the latter requires extremely efficient telescopic 
quantum time marchers far beyond the current state of the art. 
The topic is currently under active exploration. 

\begin{acknowledgments}
We thank Peter Coveney, Simona Perotto, David Spergel and Alessandro Zecchi 
for valuable discussions. CS and SS acknowledge financial support from the 
Italian National Center for HPC, Big Data and Quantum Computing (CN00000013).
SS wishes to acknowledge financial support from the Physics and Astronomy
Department of Tufts University.
\end{acknowledgments}

\end{document}